\documentclass[prl,onecolumn]{revtex4}

\usepackage{xcolor}
\usepackage[utf8]{inputenc}
\usepackage{graphicx}
\usepackage{mhchem} 
\usepackage{physunits}
\usepackage{braket}
\usepackage{float}
\usepackage{amssymb}
\usepackage{amsmath}
\usepackage{physics}
\usepackage{mhchem}
\usepackage[makeroom]{cancel}
\usepackage{upgreek}
\usepackage{commath}
\usepackage{siunitx}
\usepackage{caption,tabularx,booktabs}
\usepackage{makecell}
\usepackage{footnote}
\usepackage{chemformula} 
\usepackage[T1]{fontenc} 
\usepackage{dcolumn}
\usepackage{bm}
\usepackage[titletoc]{appendix}

\usepackage{hyperref}
\hypersetup{colorlinks=true, linkcolor=blue, citecolor=blue, urlcolor=blue}

\begin{document}
\title{Superconductivity Induced by Site-Selective Arsenic Doping in \ch{Mo5Si3}}
\author{Bin-Bin Ruan$^1$}
\email[Corresponding author:\\]{bbruan@mail.ustc.edu.cn}

\author{Jun-Nan Sun$^{2,3}$}

\author{Meng-Hu Zhou$^1$}

\author{Qing-Song Yang$^{1,4}$}

\author{Ya-Dong Gu$^{1,4}$}

\author{Gen-Fu Chen$^{1,4}$}

\author{Lei Shan$^{2,3}$}

\author{Zhi-An Ren$^{1,4,}$}
\email[Corresponding author:\\]{renzhian@iphy.ac.cn}
\affiliation{$^1$Institute of Physics and Beijing National Laboratory for Condensed Matter Physics, Chinese Academy of Sciences, Beijing 100190, China}
\affiliation{$^2$Information Materials and Intelligent Sensing Laboratory of Anhui Province, Institutes of Physical Science and Information Technology, Anhui University, Hefei 230601, China}
\affiliation{$^3$Key Laboratory of Structure and Functional Regulation of Hybrid Materials (Anhui University), Ministry of Education, Hefei 230601, China}
\affiliation{$^4$School of Physical Sciences, University of Chinese Academy of Sciences, Beijing 100049, China}

\begin{abstract}
Arsenic doping in silicides has been much less studied compared with phosphorus. 
In this study, superconductivity is successfully induced by As doping in \ch{Mo5Si3}.
The superconducting transition temperature ($T_c$) reaches 7.7 K, 
which is higher than those in previously known \ch{W5Si3}-type superconductors.
\ch{Mo5Si2As} is a type-II BCS superconductor with upper and lower critical fields of 6.65 T and 22.4 mT, respectively. 
In addition, As atoms are found to selectively take the 8\textit{h} sites in Mo$_5$Si$_2$As. 
The emergence of superconductivity is possibly due to the shift of Fermi level as a consequence of As doping, 
as revealed by the specific heat measurements and first-principles calculations. 
Our work provides not only another example of As doping, but also a practical strategy to 
achieve superconductivity in silicides through Fermi level engineering.

\end{abstract}

\maketitle

Arsenic doping in elemental silicon has been known for more than half a century.~\cite{1960As} 
And the kinetics for As diffusion 
in silicon had been well established.~\cite{1983As,1996As} 
However, reports on arsenic doping in silicides are unexpectedly scarce. 
To the best of our knowledge, there are only 5 examples in bulk materials, namely: 
Cr$_3$Si$_{1-x}$As$_x$,~\cite{1967As-doping} Fe$_5$SiAs,~\cite{1992P-doping} ZrSi$_{1-x}$As$_x$Te,~\cite{1995As-doping} and Zr(Hf)As$_{2-x}$Si$_x$.~\cite{2007As-doping,2010As-doping} 

On the other hand, phosphorus doping had been reported in more than 30 silicides: CoSi$_{0.4}$P$_{0.6}$,~\cite{1962P-doping} 
USi$_{0.17}$P$_{0.83}$,~\cite{1969P-doping} Fe$_5$SiP,~\cite{1992P-doping} 
Nb$_5$Si$_{3-x}$P$_{0.5+x}$,~\cite{2005P-doping} Gd$_5$Si$_{4-x}$P$_x$,~\cite{2009P-doping} and MSi$_x$P$_y$ (M = Fe, Co, Ru, Rh, Pd, Os, Ir, Pt, $x+y\ge4$),~\cite{1995P-doping,1997P-doping} 
to name a few. Compared to the P dopant, the much less studied As doping offers an opportunity to chase for new compounds in this field. 

Our group has been working on MoAs-based superconductors, and has discovered a series of quasi-one-dimensional superconductors $A_2$\ch{Mo3As3} ($A$ = K, Rb, Cs).~\cite{K2Mo3As3,Rb2Mo3As3,Cs2Mo3As3} 
Recently, superconductivity was reported in Re doped \ch{Mo5Si3} 
with a transition temperature ($T_c$) of 5.8 K, setting a new record in the isostructural compounds.~\cite{Mo3Re2Si3} 
We, thus, naturally conducted a systematic As doping study on \ch{Mo5Si3}, which led to the discovery of superconductivity in Mo$_5$Si$_{3-x}$As$_x$. 

\begin{figure}[h]
	\centering 
	\includegraphics{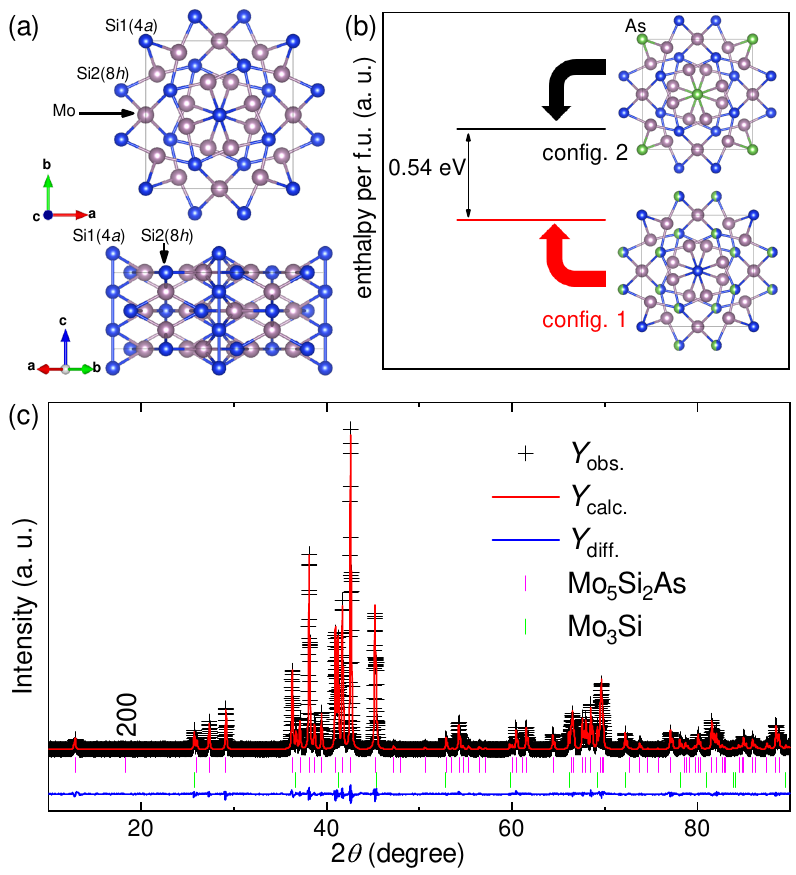} 
	\caption{(a) Crystal structure of \ch{Mo5Si3}. 
		(b) Two possible configurations of \ch{Mo5Si2As} and the corresponding enthalpies from DFT calculations. 
		(c) Room-temperature XRD pattern of \ch{Mo5Si2As} and its Rietveld refinement. 
		The vertical bars indicate the Bragg positions for \ch{Mo5Si2As} and the \ch{Mo3Si} impurity. 
		The relevant (200) peak of \ch{Mo5Si2As} is marked. }
	\label{fig:XRD}
\end{figure}

\ch{Mo5Si3} crystallizes in a tetragonal \ch{W5Si3}-type structure (space group $I4/mcm$), as illustrated in Figure~\ref{fig:XRD}(a). 
Notice there are two kind of Si atoms taking different Wyckoff positions: Si1 at the 4$a$ sites, and Si2 at the 8$h$ sites. 
In addition, Si1 atoms are bonded with each other, forming one-dimensional Si--Si chains along the $c$-axis. 
The formation of Si--Si bonds is justified by the short distance between them (2.45 \AA, which is comparable with those in $\alpha$-\ch{ThSi2}~\cite{thsi2} or \ch{CaSi2}~\cite{casi2}), 
as well as the calculated charge density (Figure \textcolor{blue}{S1}). 

In this study, we show that arsenic can be doped into \ch{Mo5Si3}. Interestingly, As atoms selectively take the Si2 (8$h$) sites. 
Furthermore, bulk superconductivity with a $T_c$ of 7.7 K is induced in \ch{Mo5Si2As}.

Polycrystalline samples of Mo$_5$Si$_{3-x}$As$_x$ (0 $\le$ $x$ $\le$ 1.25) were obtained by solid state reactions. 
Details for the preparation, characterization, and first-principles calculation can be found in the \textcolor{blue}{Supporting Information}. 
Upon As doping, the lattice parameters $a$ and $c$ increase monotonously. In our study, the doping limit for As was found to be $x$ $\sim$ 1.0. 
Superconductivity was observed in all the samples with $x$ $\ge$ 0.5, with $T_c$ increases from 4.4 K in Mo$_5$Si$_{2.5}$As$_{0.5}$ to 7.7 K in \ch{Mo5Si2As} (the optimal-doped sample). 
The data for \ch{Mo5Si2As} are shown in the main text, 
while additional information about Mo$_5$Si$_{3-x}$As$_x$ ($x$ $\ne$ 1.0) is shown in Figure \textcolor{blue}{S2--S4}. 

Figure~\ref{fig:XRD}(c) demonstrates the powder x-ray diffraction (XRD) pattern of \ch{Mo5Si2As}. 
For \ch{Mo5Si2As}, there are at least two configurations (configs.) in which As atoms take different sites. (Figure~\ref{fig:XRD}(b)) 
In config. 1, As atoms randomly take the Si2 (8$h$) sites, while in config. 2, As atoms takes all the Si1 (4$a$) sites. 
However, the observed XRD pattern can \emph{only} be refined with config. 1. 
In particular, any occupation of As at the Si1 (4$a$) sites will cause a significant enhancement of the (200) peak, which prevents the refinement from convergence. 
(See Figure \textcolor{blue}{S5})
A brief list of the refined crystallographic parameters is shown in Table~\ref{tbl:XRD}. 
More details about the refinement results can be found in Table \textcolor{blue}{S1}. 
According to the refinement, there is about 10.3 $wt.$\% of \ch{Mo3Si} impurity in the sample. 
The refined composition is Mo$_{5.00(1)}$Si$_{1.97(3)}$As$_{1.05(8)}$, which is close to that determined by energy-dispersive x-ray spectroscopy. 
(For the sample morphology and elemental mapping, see Figure \textcolor{blue}{S6}) 

\begin{table}
	\caption{Crystallographic Parameters of \ch{Mo5Si2As} from Rietveld Refinement of XRD. ($R_p$ = 1.03\%, $R_{wp}$ = 1.49\%)}
	\label{tbl:XRD}
	\resizebox{0.48\textwidth}{8mm}{
		\begin{tabular}{lccccc}
			\hline
			Atom(site)   & $x$ & $y$ & $z$ & $U_{eq}$(0.01\AA$^2$) & Occupancy \\
			\hline
			Mo1(4$b$)   & 0  & 0.5 & 0.25 & 0.137(27) & 0.987(2)\\
			Mo2(16$k$) & 0.07652(4)  & 0.22196(5)& 0 & 0.392(17) & 1.0\\
			Si1(4$a$) &  0 & 0 & 0.25 & 0.84(15)& 1.0\\
			Si2/As(8$h$) & 0.16592(8)  & 0.66592(8) & 0 & 1.36(6) & 0.474(4)/0.526 \\
			\hline
	\end{tabular}}
\end{table}

The site-selective doping of As is further backed by the first-principles calculations. 
As shown in Figure~\ref{fig:XRD}(b), the enthalpy ($H$) for config. 1 is lower compared with config. 2, 
which means As doping at the Si2 (8$h$) sites is more favorable in energy. 
The above conclusion is based on zero temperature calculations, but it is also true for finite temperatures. 
This is because the free energy $G = H-TS$, and the entropy $S$ for config. 1 is larger. 
(Notice the Boltzmann relation $S = k_B\ln \Omega$, where $k_B$ is the Boltzmann constant, and $\Omega$ is the number of microstates) 

While both the XRD refinement and first-principles calculation suggest a site-selective As doping, 
we do not rule out the possibility that a tiny amount of As atoms, which is beyond the resolution limit of powder XRD, may also takes the Si1 (4$a$) sites. 
Nevertheless, even if this is the case, the amount of As at 4$a$ sites should be negligible. 

\begin{figure}[h]
	\centering 
	\includegraphics{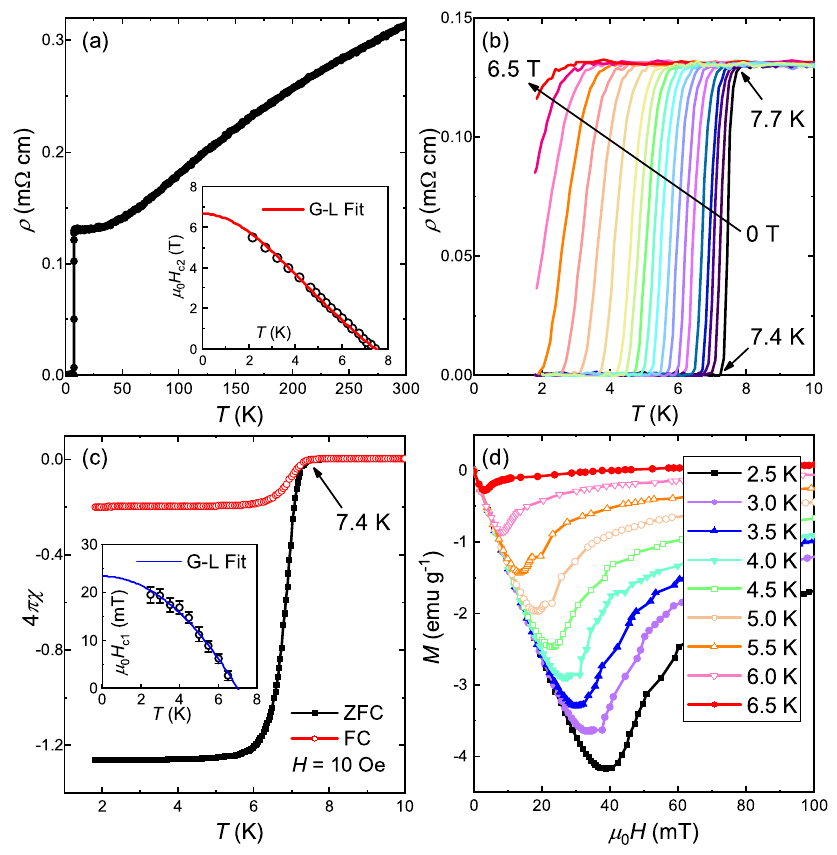} 
	\caption{(a) Temperature dependence of resistivity ($\rho$) of \ch{Mo5Si2As} under zero magnetic field. 
		Inset shows the temperature dependence of upper critical field.  
		(b) Zoom-in of the superconducting transition on $\rho(T)$ under magnetic fields from 0 T to 6.5 T. 
		(c) DC magnetic susceptibility of \ch{Mo5Si2As} under 10 Oe. 
		Inset shows the temperature dependence of lower critical field. 
		(d) Isothermal magnetization at different temperatures. }
	\label{fig:RTMT}
\end{figure}

Figure~\ref{fig:RTMT}(a) shows the temperature dependence of resistivity ($\rho$) of \ch{Mo5Si2As}. 
The monotonous decrease of $\rho$ upon cooling indicates a metallic nature. 
Superconducting transition is observed below 7.7 K ($T_c^{onset}$), with zero resistivity achieved at 7.4 K ($T_c^{zero}$). 
Figure~\ref{fig:RTMT}(b) emphasizes the region of the superconducting transition. 
Upon the application of magnetic field, the transition is gradually suppressed. 
The temperature dependence of upper critical field ($\mu_0H_{c2}(T)$) can thus be determined, which is shown in the inset of Figure~\ref{fig:RTMT}(a). 
A Ginzburg--Landau (G--L) fit gives $\mu_0H_{c2}(0)$ = 6.65(4) T. 

DC magnetic susceptibility ($4\pi\chi$) of \ch{Mo5Si2As} from 1.8 K to 10.0 K under 10 Oe is demonstrated in Figure~\ref{fig:RTMT}(c). 
Bulk superconductivity is confirmed in $4\pi\chi(T)$ curves, with large diamagnetic signals observed below 7.4 K. 
The value of $T_c$ determined from $4\pi\chi(T)$ agrees very well with that from $\rho(T)$. 
We also measured the isothermal magnetization curves for \ch{Mo5Si2As}. The results are shown in Figure~\ref{fig:RTMT}(d). 
The lower critical fields ($\mu_0H_{c1}(T)$) are determined from the deviation of the curves from initial Meissner states. 
As shown in the inset of Figure~\ref{fig:RTMT}(c), $\mu_0H_{c1}(T)$ can be fitted with the G--L relation: 
$\mu_0H_{c1}(T)=\mu_0H_{c1}(0)[1-(T/T_c)^2]$, giving $\mu_0H_{c1}(0)$ = 22.4(5) mT. 

A series of superconducting parameters can be determined using $\mu_0H_{c2}(0)$ and $\mu_0H_{c1}(0)$. 
The values of these parameters are listed in Table~\ref{tbl:SCpara}, which includes the G--L coherence length ($\xi_{GL}$),
the penetration depth ($\lambda_{GL}$), the G--L parameter ($\kappa_{GL}$), and the thermodynamic field ($H_{c}(0)$). 
Details for calculation of the parameters can be found in the \textcolor{blue}{Supporting Information}. 
Notice $\kappa_{GL} \gg 1/\sqrt{2}$, suggesting type-II superconductivity in \ch{Mo5Si2As}.

\begin{figure}[h]
	\centering 
	\includegraphics{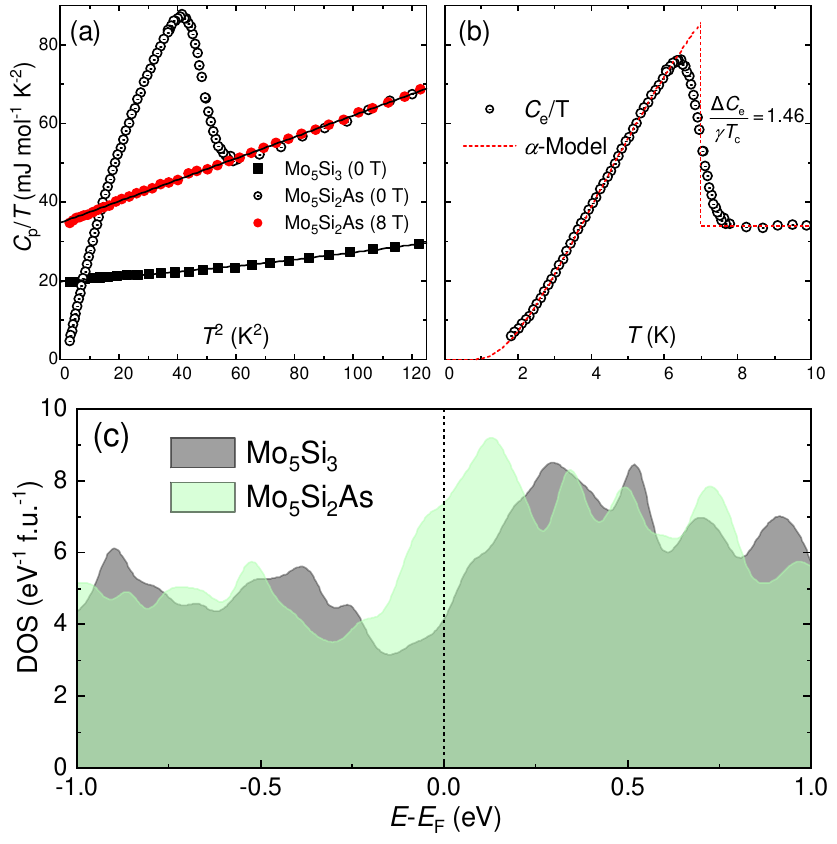} 
	\caption{(a) Temperature dependence of specific heat ($C_p$) of \ch{Mo5Si2As} under zero magnetic field, and under a field of 8 T. 
		The $C_p$ data for \ch{Mo5Si3} are also plotted for comparison. Solid lines are fittings with Debye model. 
		(b) Temperature dependence of the electronic contribution of $C_p$ under zero magnetic field. (a) and (b) share the same $y$-axis range. 
		(c) Calculated DOS for \ch{Mo5Si3} and \ch{Mo5Si2As} near the Fermi level. }
	\label{fig:Cp}
\end{figure}

The specific heat ($C_p$) of \ch{Mo5Si2As} was measured to investigate the superconducting, as well as the thermodynamic properties. 
The results are shown in Figure~\ref{fig:Cp}(a). Notice the data have been corrected to exclude the influence of \ch{Mo3Si} impurity. 
Thermodynamic parameters for \ch{Mo3Si} are adopted from the literature.~\cite{Mo3Si} 
Under zero magnetic field, $C_p$ for \ch{Mo5Si2As} clearly shows an anomaly at $\sim$ 7.0 K, validating the bulk superconductivity. 
When a field of 8 T was applied, superconductivity was completely suppressed. The data under 8 T can be well fitted with the Debye model: 
$C_p(T) = \gamma T + \beta T^3$, giving $\gamma$ = 34.74(8) mJ mol$^{-1}$ K$^{-2}$, and $\beta$ = 0.273(1) mJ mol$^{-1}$ K$^{-4}$. 
The Debye temperature ($\Theta_D$) is thus determined by $\Theta_D = (12\pi^4NR/5\beta)^{1/3}$ to be 385(2) K. 
($N$ is the number of atoms in a formula unit ($f.u.$), and $R$ is the ideal gas constant) 
The electronic contribution of specific heat ($C_e$) is extracted by: $C_e = C_p|_{\mu_0H=0}-C_p|_{\mu_0H=8 T}+\gamma$, 
which is shown in Figure~\ref{fig:Cp}(b). $C_e$ at the superconducting state can be well fitted with the so-called $\alpha$-model.~\cite{alphaModel} 
The normalized $C_e$ change $\Delta C_e/\gamma T_c$ = 1.46(2), which is close to the BCS weak-coupling ratio (1.43). 
The electron-phonon coupling strength ($\lambda_{ep}$) can be estimated by using the inverted McMillan formula:~\cite{McMillan1968} 
\begin{equation}
	\label{eq:McMillan}
	\lambda_{ep} = \frac{1.04+\mu^*\ln(\Theta_D/1.45T_c)}{(1-0.62\mu^*)\ln(\Theta_D/1.45T_c)-1.04}.
\end{equation}
By setting the Coulomb screening parameter $\mu^*$ = 0.13, a typical value for intermetallics, we get $\lambda_{ep}$ = 0.66(1). 
These results suggest a weak to moderate coupling in \ch{Mo5Si2As}. 

For comparison, we also measured the temperature dependence of $C_p$ of \ch{Mo5Si3}, which is shown in Figure~\ref{fig:Cp}(a). 
$\gamma$ and $\Theta_D$ for \ch{Mo5Si3} are estimated to be 19.80 mJ mol$^{-1}$ K$^{-2}$ and 659 K, respectively. 
These values are comparable with previous report on \ch{Mo5Si3} single crystals.~\cite{Mo5Si3sing} 
Notice that $\gamma$ in \ch{Mo5Si2As} is significantly larger than the undoped \ch{Mo5Si3}, 
and the large decrease of $\Theta_D$ suggests substantial softening of the lattice. 
To take more insight into this, the density of states (DOS) was calculated for \ch{Mo5Si3} and \ch{Mo5Si2As}. 
The results are shown in Figure~\ref{fig:Cp}(c). One may notice the similar shapes of the DOS curves, which means that the bands can be considered rigid in our case. 
(For the band structures, see Figure \textcolor{blue}{S7}) 
Arsenic doping shifts the Fermi level ($E_F$) to higher energy, causing a significant enhancement of DOS at $E_F$ ($N(E_F)$). 
This is as expected, since As hosts one more valence electron compared with Si. 
$N(E_F)$ for \ch{Mo5Si3} is 4.13 $eV^{-1} f.u.^{-1}$, while $N(E_F)$ for \ch{Mo5Si2As} is 7.37 $eV^{-1} f.u.^{-1}$. 
The enhancement of $N(E_F)$ lead to the change of $\gamma$. We can theoretically estimate the value of $\gamma$ for \ch{Mo5Si2As} by: 
\begin{equation}
	\label{eq:gamma}
	\gamma = \frac{1}{3}N(E_F)\pi^2k_B^2(1+\lambda_{ep})
\end{equation}
to be $\sim$ 29.02 mJ mol$^{-1}$ K$^{-2}$, which agrees well with the experimental value. 
According to McMillan's formalism, $\lambda_{ep} = [N(E_F)\langle I^2\rangle ]/[M\langle \omega^2\rangle ]$, 
where $M$ is the atomic mass, $\langle I^2\rangle $ and $\langle \omega^2\rangle $ stand for averages of the squared electronic matrix elements on the Fermi surface, 
and of the squared phonon frequencies, respectively.~\cite{McMillan1968}  
The emergence of superconductivity in Mo$_5$Si$_{3-x}$As$_x$ should be due to the enhancement of $N(E_F)$, 
and possibly the softening of lattice (as evidenced by the large decrease of $\Theta_D$). 

\begin{table}
	\caption{Superconducting and Thermodynamic Parameters of \ch{Mo5Si2As}.}
	\label{tbl:SCpara}
	\begin{tabular}{lc}
		\hline
		Parameter (unit)   & Value \\
		\hline
		$T_c^{onset}$ (K)   & 7.7(1)  \\
		$T_c^{zero}$ (K) & 7.4(1) \\
		$\mu_0H_{c1}(0)$ (mT) & 22.4(5) \\
		$\mu_0H_{c2}(0)$ (T) & 6.65(4) \\
		$\mu_0H_{c}(0)$ (T) & 0.22(0) \\
		$\xi_{GL}$ (nm) & 7.03(2) \\
		$\lambda_{GL}$ (nm) & 159.5(4) \\
		$\kappa_{GL}$ & 22.7(1) \\
		$\gamma$ (mJ mol$^{-1}$ K$^{-2}$) & 34.74(8) \\
		$\beta$ (mJ mol$^{-1}$ K$^{-4}$) & 0.273(1) \\
		$\Theta_D$ (K) & 385(2) \\
		$\lambda_{ep}$ & 0.66(1) \\
		$\Delta C_e/\gamma T_c$ & 1.46(2) \\
		\hline
	\end{tabular}
\end{table}

\begin{figure}[th]
	\centering 
	\includegraphics{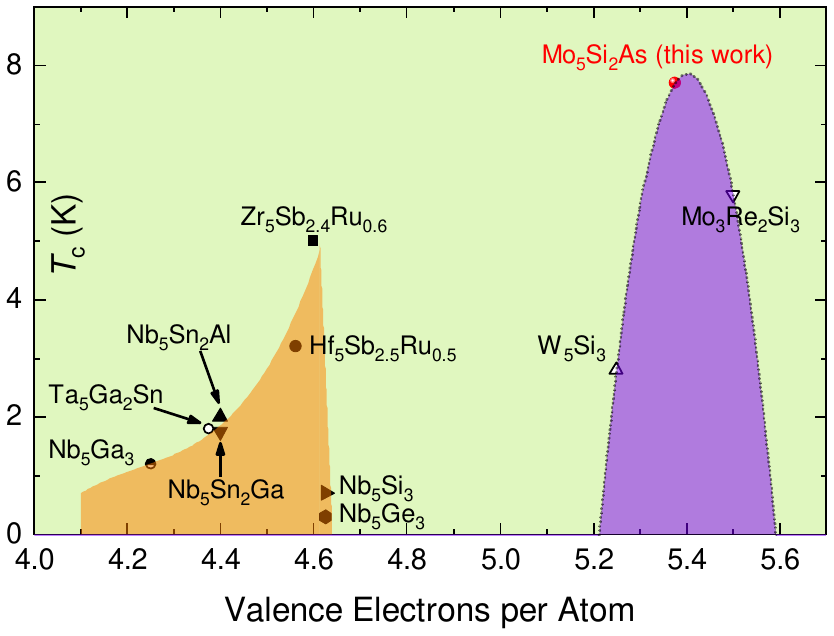} 
	\caption{Evolution of $T_c$ in \ch{W5Si3}-type superconductors upon the valence electron per atom ($e/a$). 
		Notice the peak at $e/a \sim 5.4$ is just a tentative guide since there are less examples with $e/a > 5$. }
	\label{fig:Tc}
\end{figure}

Last but not least, we would like to point out that a $T_c$ of 7.7 K is fairly high in \ch{W5Si3}-type superconductors. 
A comparison of $T_c$ between \ch{Mo5Si2As} and previously reported \ch{W5Si3}-type superconductors 
is illustrated in Figure~\ref{fig:Tc}.~\cite{Nb5Ga3,Nb5Sn2Ga,Ta5Ga2Sn,Nb5Ge3,Zr5Sb3,Hf5Sb3,Nb5Sn2Al,W5Si3,Mo3Re2Si3}  
One may notice that the average number of valence electrons per atom ($e/a$) clearly affects the value of $T_c$ in this structural family. 
Maximal $T_c$ occurs when $e/a \sim$ 4.6 or 5.4. Notice the second peak (at $e/a \sim 5.4$) is just a tentative guide, 
and is less reliable as there are only a few examples with $e/a > 5$. 
The phenomenon that $T_c$ peaks at certain $e/a$ values, known as the Matthias rule, has also been observed in other intermetallics.~\cite{MatthiasRule} 
Currently, the region of $e/a > 5.5$ remains unexplored, more superconductors may be discovered if more electrons can be introduced into this structural family.  

In summary, we report the discovery of superconductivity in Mo$_5$Si$_{3-x}$As$_x$ (0.5 $\le$ $x$ $\le$ 1.25), 
in which a maximal $T_c$ of 7.7 K is observed in \ch{Mo5Si2As}. 
Arsenic doping in \ch{Mo5Si3} is found to be site-selective. 
According to specific heat measurements and first-principles calculations, 
the emergence of superconductivity is related to the shift of Fermi level. 
Our method of As doping should be generally applicable to other silicides, in which more superconductors can be expected. 

\textbf{Note}: \textit{This manuscript has been published in June, 2022. We also reported superconductivity in Mo$_5$Si$_{3-x}$P$_x$ (maximal $T_c$ = 10.8 K), see arXiv: 2208.02392}

\section{Conflict of interest}
The authors declare that they have no conflict of interest.

\section{Acknowledgments}
This work was supported by the National Key Research and Development of China (Grant Nos. 2018YFA0704200, 2021YFA1401800, 2018YFA0305602, and 2017YFA0302904), 
the National Natural Science Foundation of China (Grant Nos. 12074414, 12074002, and 11774402), 
and the Strategic Priority Research Program of Chinese Academy of Sciences (Grant No. XDB25000000).

\bibliography{MSA_bib}

\end{document}